\documentclass[preprint2]{aastex} 

\slugcomment{Submitted to ApJ}

\shorttitle{High redshift 3CR sources}
\shortauthors{Haas et al.}

\begin{document}


\title{Near- and mid-infrared photometry of high-redshift 3CR sources 
}


\author{Martin Haas\altaffilmark{1}, S. P. Willner\altaffilmark{2},
  Frank Heymann\altaffilmark{1,3},
  M. L. N. Ashby\altaffilmark{2}, \\
  G. G. Fazio\altaffilmark{2}, Belinda J. Wilkes\altaffilmark{2}, 
  Rolf Chini\altaffilmark{1}, and Ralf Siebenmorgen\altaffilmark{3}
}

\altaffiltext{1}{Astronomisches Institut, Ruhr-Universit\"at Bochum,
              Universit\"atsstra{\ss}e 150, 44801 Bochum, Germany,\\
              email: haas@astro.rub.de}
\altaffiltext{2}{Harvard-Smithsonian Center for Astrophysics, \\
  60 Garden Street, Cambridge MA 02138, USA}
\altaffiltext{3}{European Southern Observatory, \\
  Karl-Schwarzschildstr. 2, 
             85748 Garching, Germany}


\begin{abstract}
  Using the Spitzer Space Telescope, we have obtained 
  3.6--24\,$\mu$m photometry of 38 radio galaxies and 24
  quasars from the 3CR catalog at redshift $1<z<2.5$. 
  This 178~MHz-selected sample is 
  unbiased with respect to orientation and therefore 
  suited to study orientation-dependent effects 
  in the most powerful active galactic nuclei (AGN).  Quasar and
  radio galaxy subsamples matched in isotropic radio luminosity are compared.
  The quasars all have similar spectral 
  energy distributions (SEDs), nearly constant 
in $\nu$\,F$_{\nu}$
through the rest
  1.6--10~\micron\ range, consistent with a 
  centrally heated dust distribution which outshines 
  the host galaxy contribution. 
  The radio galaxy SEDs show larger dispersion, but
  the mean radio galaxy SED declines from rest 1.6 to 3\,$\mu$m and then
  rises from 3 to 8\,$\mu$m.    
  The radio galaxies are on average a factor 3--10 less
  luminous in this spectral range than the quasars.  These
  characteristics are  consistent with composite
  emission from a heavily reddened AGN plus starlight from the host
  galaxy.
  The mid-infrared colors and radio to mid-infrared spectral slopes
  of individual galaxies are also 
  consistent with this picture.  Individual galaxies show different
  amounts of extinction   and host galaxy starlight,
  consistent with the orientation-dependent unified scheme. 

\end{abstract}





\section{Introduction}

When exploring the general evolution of galaxy populations
across cosmic times,  
a particular challenge is to distinguish between black hole
and star-forming activity. 
Star formation and obscuring dust go hand in hand, and 
black-hole-driven active galactic nuclei (AGN) are also surrounded by dust mainly 
distributed in a disk/torus-like geometry (Antonucci 1993).
There is evidence that AGN mainly power the near- and mid-IR emission
(NIR, $\sim$2\,$\mu$m; MIR, $\sim$10\,$\mu$m)
from hot nuclear dust, 
while starbursts contribute mainly to the 
far-infrared  (FIR, $\sim$60\,$\mu$m) luminosity
(e.g., Rowan-Robinson 1995, 
Vernet et al.\ 2001, 
Schweitzer et al.\ 2006).
Using the MIR/FIR luminosity ratio as an indicator
for the relative AGN and starburst contributions, numerous studies
have found an increase of AGN/starburst activity with total luminosity and
redshift, but the validity of this trend is still under
discussion because of selection effects on the various samples.
More seriously,  
an unfavorable AGN orientation could cause MIR
obscuration (e.g. Pier \& Krolik 1993), leading to a fundamental 
observational degeneracy:
a low MIR/FIR luminosity ratio can be due to either a high star-forming
contribution or to an AGN in which the hot dust is obscured.
The spectral energy distribution (SED) of an obscured AGN may thus mimic
that of a starburst-powered source. 
While this degeneracy has now been widely examined  
at low/intermediate luminosity and redshift ($z<1$), it has still to
be explored for the most luminous sources at high redshift ($z>1$).
In order to assess galaxy and AGN evolution 
in the universe, we therefore need to understand this AGN/starburst  
degeneracy for a population of luminous high-redshift sources. 
A crucial step towards this is to study the orientation dependence of
the NIR and MIR emission of high-redshift AGN.

Orientation-dependent effects  
can only be tested and quantified with
AGN samples having type~1 (unobscured) and type~2 (obscured)
subsamples matched in isotropic emission. 
The clean AGN tracers ---
optical, [\ion{O}{3}]~$\lambda$5007\,\AA, NIR, and X-ray ($\la$10~keV) 
--- all fail to fulfill this requirement.  
The [\ion{O}{2}]~$\lambda$3727\AA ~emission, while
isotropic (Hes et al.\ 1993), is probably dominated by extended 
starbursts and shocks (Best et al.\ 2000) rather than by the AGN. 
Therefore, the only feasible way is low-frequency
(meter-wavelength) radio selection because 
the integrated emission from the radio lobes is optically thin
and essentially isotropic.
This makes radio-loud AGN particularly attractive for studying
orientation-dependent properties at other wavelengths and, 
after sorting out the influence of radio jets/lobes on the emission, 
for generalizing conclusions about orientation-dependent effects 
to the much larger population of radio-quiet AGN. 

The brightest low-frequency-selected AGN sample is the 
3CR compilation (Spinrad et al.\ 1985). 
The powerful double-lobed radio galaxies
(henceforth simply called radio galaxies) are supposed to be
misaligned quasars (Barthel 1989). 
Based on {\it IRAS} coadded scans and a few individual detections, 
Heckman et al.\ (1992, 1994) already noted an average MIR/FIR difference 
between 3CR quasars and radio galaxies. More comprehensive  
MIR and FIR spectrophotometry from {\it ISO} is in hand 
(as compiled by Siebenmorgen et al.\ 2004 and by 
Haas et al.\ 2004) as well as from {\it Spitzer} (e.g., Shi et al.\ 2005, 
Haas et al.\ 2005, Ogle et al.\ 2006, Cleary et al.\ 2007), providing
a basis to study the $z < 1$ 3CR objects.  
These sources are, however, a factor of five less radio-luminous on
average than the most powerful radio sources seen at 
higher redshift, and the lower indicated accretion power may reflect
different source physics.
The higher-luminosity population can be sampled by
the 3CR sources at  $1<z<2.5$, which have radio luminosities similar
to those of 
the most powerful radio sources at even higher redshift ($2.5<z<6$).
However, with the exception of a few targets 
(Siebenmorgen et al.\ 2004, Seymour et al.\ 2007), 
the high-$z$ 3CR sample has not been well observed 
in the rest frame NIR and MIR.  
 
This paper is based on {\it Spitzer} observations of 62 of the 64
high-$z$ 3CR sources.  
It focuses on the hot nuclear dust emission and its obscuration
in the most luminous type-2 AGN.
We use a $\Lambda$CDM cosmology with
$H_0 =71$~km/s/Mpc, $\Omega_{{\rm m}} = 0.27$
and $\Omega_{\Lambda} = 0.73$.

 \begin{figure*}[htb]
   \begin{center}
   \includegraphics[height=14.5cm, angle=90]{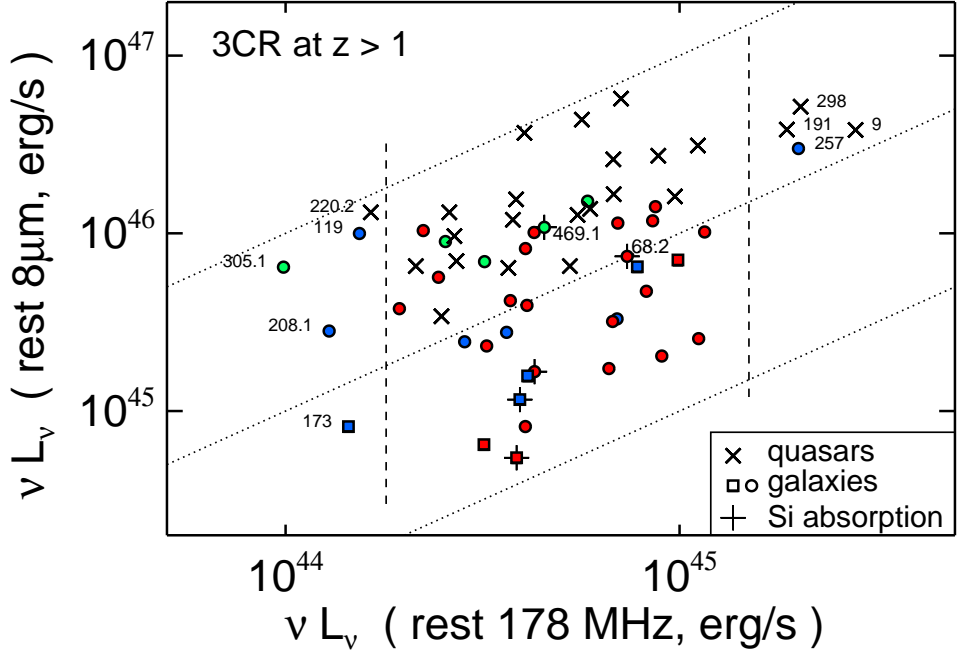}
   \caption{Infrared versus radio luminosity of the 3CR sample at 
     $z>1$ prior to normalization.  'x' symbols denote quasars;
     circles and squares denote radio galaxies.
     Superposed crosses indicate radio galaxies that show evidence of
   silicate absorption (\S\ref{sec_rg}).
     The vertical long-dashed lines mark the range of our
   luminosity-matched  quasar and radio galaxy subsamples. 
    The dotted lines indicate $L_{8\micron}/L_{\rm 178\,MHz}$ 
     ratios of 1, 10, and 100. 
     The radio galaxies are grouped into several SED classes 
     in Fig.\,\ref{fig_nir_mir_cc} and \S\ref{sec_rg}.  
     The color-coding and symbols are:  green circle (A), red circle (B),
     red square (C), blue square (D), blue circle (E). 
      The two low-excitation radio galaxies 3C\,68.1 and 3C\,469.1
     are labeled with their 3C numbers, as are sources outside the
   luminosity range of our analysis.
      }
   \label{fig_ir_radio_lum}
   \end{center}
 \end{figure*}

\section{Observations and Data}


With the {\it Spitzer Space Telescope} (Werner et al.\ 2004), we are
observing the entire sample of 64 high-$z$ 3CR sources 
using the instruments IRAC (3.6--8.0\,$\mu$m, Fazio et al.\ 2004), 
the IRS blue peak-up array  (16\,$\mu$m, Houck et al.\ 2004) and MIPS 
(24\,$\mu$m, Rieke et al.\ 2004). 
Most observations are performed in our guaranteed time program
(PID 40072; PI G.\ Fazio) with on-source exposure times 
4$\times$30\,s (each IRAC band), 4$\times$14\,s (IRS), 
and 10$\times$10\,s (MIPS). 
A few sources have been observed in other programs,
and we use the published photometry if available 
(e.g., PID 3329; PI D.\ Stern, Seymour et al.\ 2007). 

For IRAC, we used the basic calibrated data products (BCD, version S16) and
coadded them to 0$\farcs$869 pixels using the latest version of
{\tt IRACProc} (Schuster et al.\ 2006). This optimally handles the
slightly undersampled IRAC PSF in order to assure 
accurate point-source photometry.
For IRS, we used the post-BCD pipeline product, version S16. 
For MIPS, we used custom routines to modify
the version S17 BCD files to remove instrumental artifacts (e.g., residual
images) before shifting and coadding to create the final mosaics.
All sources are well seen on the images in all filters. 
The sources were extracted and matched using the SExtractor tool 
(Bertin \& Arnouts 1996). 
We used sufficiently large 
apertures so that aperture corrections are small ($<$5\%).
The photometric errors are typically smaller than 10\% but increase
for faint sources; 
exceptions are 3C\,225A and 3C\,294, where nearby bright stars make
the photometry uncertain in the shorter IRAC bands.

As of 2008 April, 24 quasars and 38 radio galaxies have been 
observed, covering the complete
high-$z$ 3CR sample with the exception of the quasar 3C\,245
and the radio galaxy 3C\,325. 
All 62 sources have IRAC measurements and are observed in at least 
one of the 16 and 24\,$\mu$m bands (54 sources at 16\,$\mu$m and 
60 sources at 24\,$\mu$m). 

For the analysis it is desirable to compare rest frame SEDs with the
same wavelength sampling. 
Depending on the redshift of our sources ($1<z<2.5$) 
the observations sample different rest wavelengths between 1.6 and 10\,$\mu$m.
Before resampling and interpolating the SEDs, we checked
that spectral features do not affect the interpolation. 
In principle, 
prominent spectral features in this wavelength range could be PAH emission
bands around 7.7\,$\mu$m and the 9.7\,$\mu$m silicate absorption
feature. IRS spectra
of several low-$z$ and even a few high-$z$ FR\,II radio sources are available.
PAH features are weak and usually undetected, and 
the continua are generally smooth.
Strong silicate absorption is, however, present in some objects
(Haas et al.\ 2005, Ogle et al.\ 2006, Cleary et al.\ 2007, Seymour et al.\ 2008). 
Our broadband SEDs therefore represent the
smooth continua for $\lambda\la8$~\micron\ but are uncertain at rest
wavelengths near 10~\micron.  In practice,
the de-redshifted SEDs were  interpolated in log-log space at 12
wavelengths between rest 1.6 
and 10\,$\mu$m to produce the figures shown.
 
 \begin{figure*}[htb]
   \begin{center}
   \includegraphics[height=14.5cm,angle=90,clip=true]{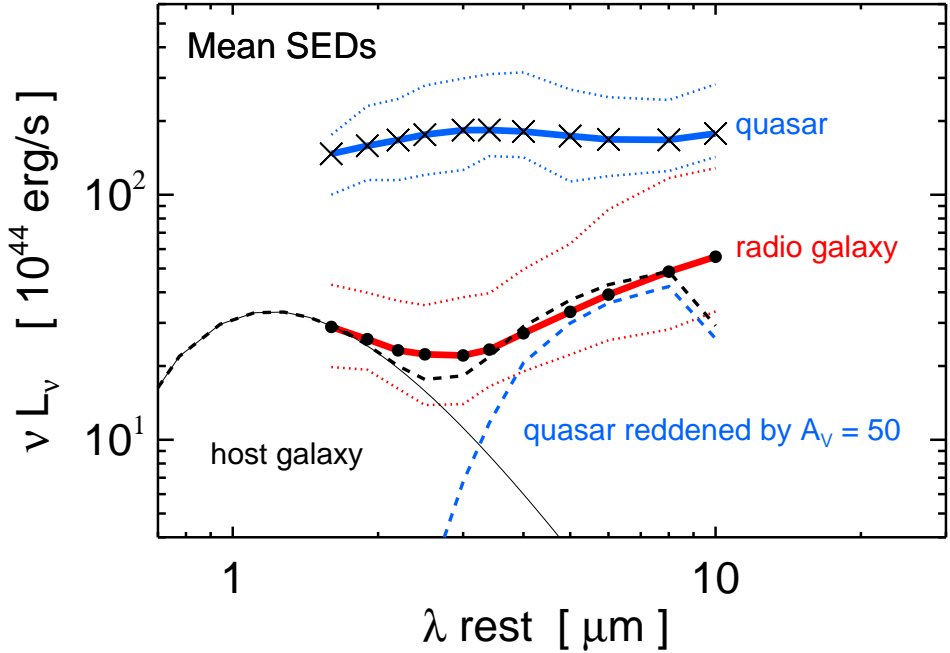}
   \caption{Rest frame quasar and radio galaxy SEDs  normalized
   by rest 178\,MHz luminosity ($\langle L_{178}\rangle$ = 5.4$\times$10$^{\rm 44}$ erg/s).
     Symbols connected with thick blue and red lines show the mean
   SEDs for quasars and radio galaxies, respectively. 
     The thin dotted blue and red lines indicate the dispersion 
     (upper and lower quartiles) around the mean SEDs; the mean
     ratio of upper/lower quartiles are 2.3 (quasars) and 3.4 (radio galaxies).
     The radio galaxy SED can be explained by the sum 
     (black long-dashed line) of a
     reddened quasar (blue long-dashed line)
     and starlight from the host galaxy (thin black solid line).
     The long-dashed lines have been shifted slightly
     to make them visible in the plot.
     The difference between radio galaxies and reddened quasars
     at 10\,$\mu$m may be due to the 
     silicate absorption feature which may escape detection in our broad
     band photometry.
   }
   \label{fig_qso_gal_seds}
   \end{center}
 \end{figure*}

The quasar and radio galaxy samples match reasonably well in redshift and 
rest-frame 178\,MHz flux density.  Rest-frame 178\,MHz radio flux
densities were derived from 
data listed in the NASA Extragalactic Database, NED.
The quasars have mean redshift $\langle z\rangle = 1.44\pm0.31$
and mean flux density $\langle S_{178}\rangle = 27.8\pm15.1$~Jy, while the
values for the radio galaxies are
$\langle z\rangle = 1.42\pm0.31$, 
$\langle S_{178}\rangle = 22.2\pm6.2$~Jy. 
Thus over the whole sample,
quasars are  about 30\% more luminous than radio 
galaxies as shown in
Figure~\ref{fig_ir_radio_lum}. 
In order to improve the luminosity match, we have excluded the sources at the 
low and high ends of the distribution,
$L_{178}<1.8\times10^{44}$\,erg/s and 
$L_{178}>1.5\times10^{45}$\,erg/s, respectively.
We have also excluded the quasar 3C\,418 
because of its flat radio spectrum 
(low-frequency spectral index $\alpha_{178}\approx0$), while all
other sources 
have steep radio spectra ($-1.1\la \alpha_{178} \la -0.6$).
The resulting mean radio luminosities of the sample galaxies are
$\langle L_{178}\rangle = (5.35\pm2.53)\times 10^{44}$~erg/s for
quasars and   
$(5.55\pm2.34)\times 10^{44}$ for radio galaxies.

While the quasar and radio galaxy distributions match very well in $L_{178}$, 
a proper analysis of orientation-dependent effects requires also 
that the individual SEDs are normalized by the radio
luminosity, which serves as a tracer for the intrinsic AGN strength. Therefore, 
we have normalized each SED to the sample mean 178\,MHz luminosity; 
after normalization each object has 
$L_{178} = 5.4\times10^{44}$~erg/s. 
Because of the good $L_{178}$ match of the samples, it turned out that 
the net effect of this normalization on the results is small. 






\section{Results and Discussion}






\subsection{Radio galaxies as obscured quasars}
\label{sec_rg}
The NIR--MIR SEDs of quasars are all very similar in shape, 
as shown in Figure~\ref{fig_qso_gal_seds}.
The SEDs can be described by a single power law 
$L_\nu \propto \nu^{- 1}$,  consistent with previous results for
lower-redshift objects (e.g., Elvis et al.\ 1994).
The dispersion of the  SEDs 
is essentially caused by differing ratios of MIR to radio luminosity.    
Some quasars exhibit small (10-20\%) bumps around 5\,$\mu$m 
explainable by distinct hot dust components.\footnote{Despite the
 similarity of the infrared SEDs, the quasar
 population is not homogeneous at optical wavelengths: 
 there are quasars like 3C\,186 with blue
 optical SED and 3C\,68.1 with red optical SED, as listed in NED.
 In the orientation-based unified scheme, 3C\,68.1 could be borderline
 so that the broad lines are detected, but most of the UV-optical continuum 
 is absorbed.}  
The power law shape of the quasar SED can naturally be explained 
by the superposition of centrally-heated dust components with 
a radial temperature gradient (1500\,K~$>T>300$\,K)
as has been found also in lower luminosity type-1 AGN 
(e.g.\ Ward et al.\ 1987, Barvainis 1987; see also Rowan-Robinson
1980). 
Any contribution of the quasar host galaxies to the NIR--MIR SEDs appears 
to be outshone (factor $\ga$5--10) by the AGN dust emission. 

In contrast to quasars, radio galaxies display a diversity of SED shapes
leading  to a 50\% larger dispersion around their mean SED
(Fig.~\ref{fig_qso_gal_seds}).
Despite the dispersion, nearly all radio galaxy SEDs show a decline from
rest 1.6\,$\mu$m to 3\,$\mu$m 
and a rise from 3\,$\mu$m to 8\,$\mu$m
($L_\nu \propto \nu^{- 1.9}$). 
In addition, the average radio galaxy SED is fainter by a factor of
three at 8\,$\mu$m 
and a factor of eight at 2\,$\mu$m relative to the quasar SED. 
Unlike the quasars, hot ($T>750$\,K) 
dust emission is not seen in the radio galaxy SEDs.  Its absence can
be explained by  
absorption (screen $A_{V}\approx50$)\footnote{ 
  The reddening curve used is a compromise
  between the latest results for Milky Way reddening and earlier data
  (summarized by Indebetouw et al.\ 2005): 
   A$_{V}$\,/\,A$_{H}$\,/\,A$_{\rm 3\mu m}$\,/\,A$_{\rm 5\mu
     m}$\,/\,A$_{\rm 8\mu m}$\,/\,A$_{\rm 10\mu m}$ = 
   1\,/\,0.184\,/\,0.070\,/\,0.037\,/\,0.028\,/\,0.040.
   }   
of the central dust emission.  The short wavelength
($\lambda< 3$~\micron) component can then be explained by emission from
stars in the host galaxy. 
Extrapolation of the mean 3--8\,$\mu$m SED slopes 
towards longer wavelengths suggests that the radio galaxy and quasar 
SEDs meet each other at about 25--40\,$\mu$m, and  beyond these wavelengths
extinction may be no longer relevant. 

 \begin{figure*}[htb]
   \begin{center}
   \includegraphics[height=14.5cm, angle=90]{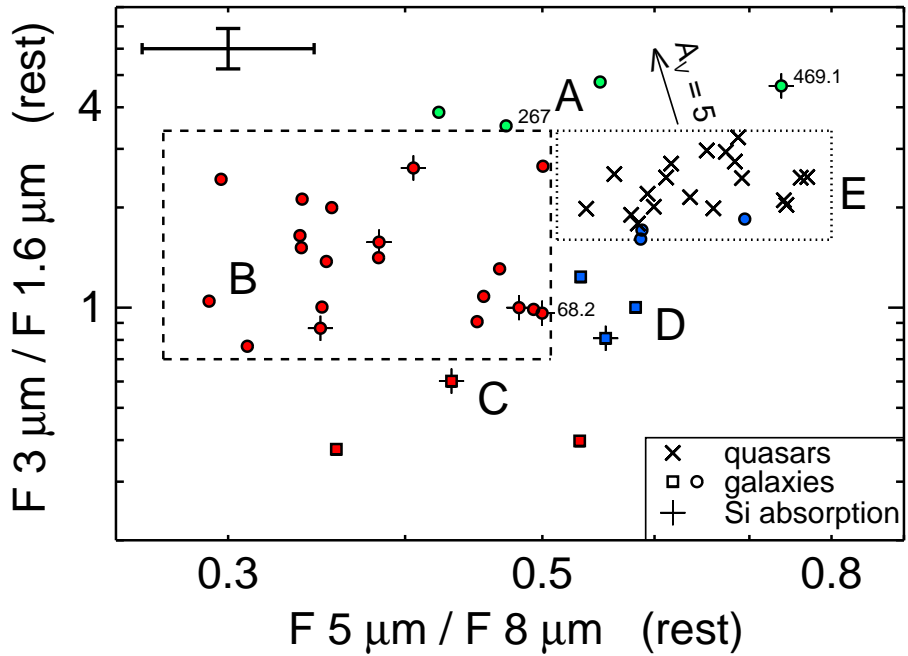}
   \caption{NIR/MIR color-color diagram. 
     The radio galaxies are grouped into five classes labeled A--E as
   explained in \S\ref{sec_rg}.  Symbol color-coding is the same as
   in Figure~\ref{fig_ir_radio_lum}.
     Radio galaxies with a photometric 
     signature for silicate absorption are additionally marked with an
     underlying plus. 
     The two sources 3C\,267 and 3C\,469.1 with 
     spectroscopically-detected silicate 
     absorption are labeled, as are the two low-excitation radio 
     galaxies 3C\,68.1 and 3C\,469.1. 
     The error bar in the upper left corner represents a color rms of 15\%.
     The $A_{V}$ arrow indicates screen extinction with 
     the reddening law given in \S\ref{sec_rg}.
}
   \label{fig_nir_mir_cc}
   \end{center}
 \end{figure*}

 \begin{figure*}[htb]
   \begin{center}
   \includegraphics[height=14.5cm,angle=90,clip=true]{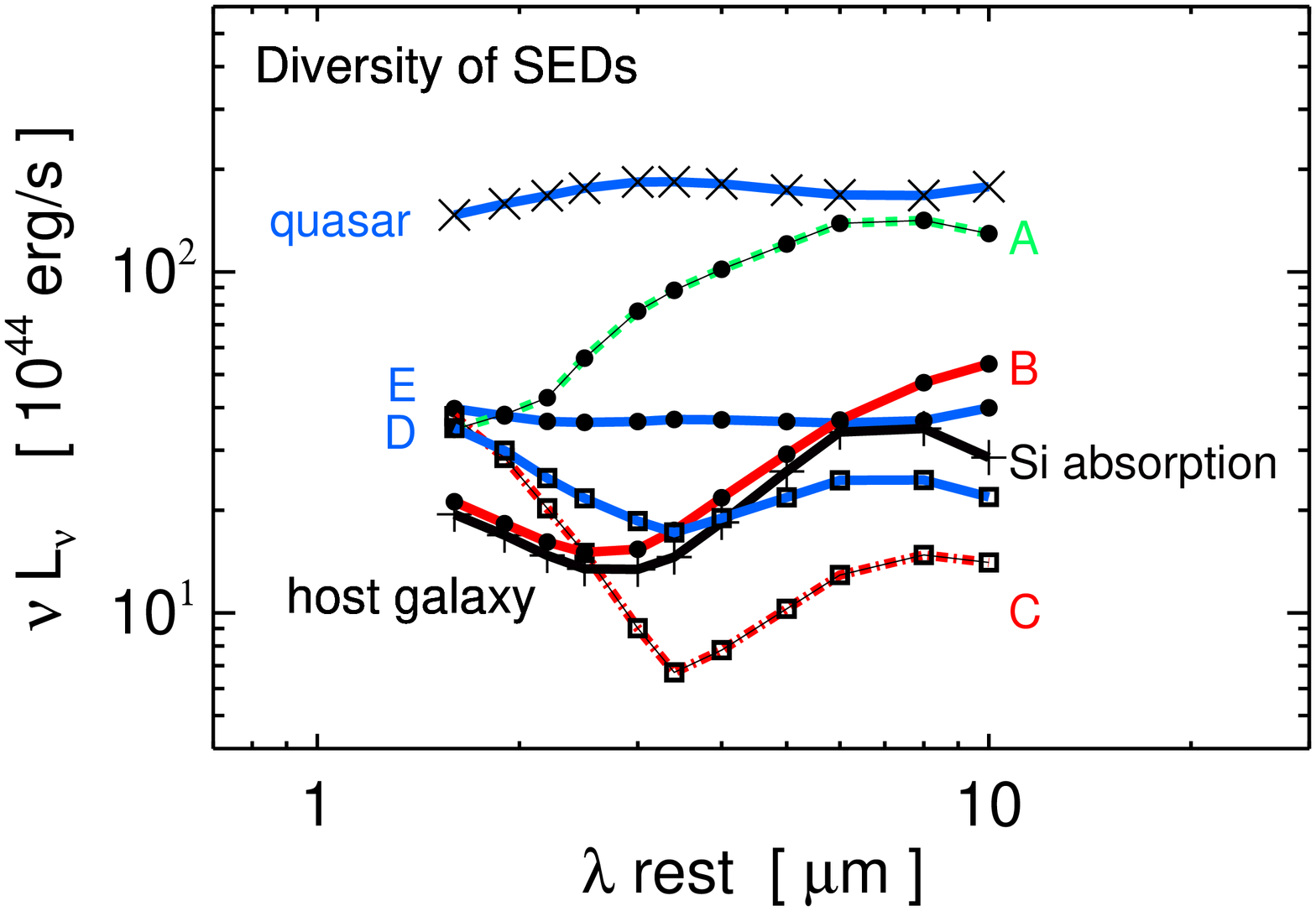}
   \caption{ 
     Mean SED of each radio galaxy class as identified in \S\ref{sec_rg}.
     The SEDs have been normalized to the mean 178\,MHz luminosity.  
     The mean quasar SED is also shown for comparison. 
     The dispersion around each SED 
     (measured as mean ratio of upper/lower quartile) is: 
     2.3 (quasars), 2.0 (A), 3.2 (B), 2.4 (C), 1.7 (D), 1.5 (E), and 4.8
   (silicate absorption).
   }
   \label{fig_gal_seds}
   \end{center}
 \end{figure*}

As noted above, the quasar NIR--MIR SED shapes are extremely
homogeneous. This
is reflected 
in the narrow range of the quasars' NIR and MIR colors.
The color-color diagram shown in Figure~\ref{fig_nir_mir_cc}
illustrates the differing SED types.
In this diagram, quasars populate a distinct locus (``E''),
while radio galaxies show wider dispersion as mentioned above. 
According to their location in the color-color diagram, we have grouped 
the radio galaxies into five classes described below.
Their SED shapes are illustrated in Figure\,\ref{fig_gal_seds}. 

\begin{itemize}
\item[A)] Four sources at the high end of the 3$\mu$m/1.6$\mu$m ratio
   (above the dotted and dashed boxes in 
   Fig.\,\ref{fig_nir_mir_cc}): 
   basically, they have quasar-like SEDs, but the hottest dust
   emission at about 1--2\,$\mu$m appears to be absorbed 
   (screen A$_{\rm V}$\,$\approx$\,5) leading to a
   redder 3\,$\mu$m/1.6\,$\mu$m color compared to quasars.

\item[B)] The bulk of radio galaxies (20 sources)
   have  declining 1--3\,$\mu$m SEDs with a steep 3--8\,$\mu$m rise. 
   Their
   colors can be explained by a heavily reddened AGN plus an added
   component of starlight contributing at 1.6\,$\mu$m.
   If this explanation is correct,
   the direction of the extinction vector A$_{\rm V}$ becomes 
   meaningless
   because host galaxy starlight will not be affected by extinction
   near the nucleus.  Instead,
   vertical position in the plot is determined by the relative
   contributions of starlight and AGN light, while horizontal
   position measures the amount of extinction (to the extent the
   underlying AGN SEDs are the same).  As noted above, $A_V \sim 50$~mag
   is required to explain the colors.

\item[C)] Three sources at the low end of the 3\,$\mu$m/1.6\,$\mu$m ratio 
   (below the long dashed line in  Fig.\,\ref{fig_nir_mir_cc}): 
   Their SEDs show a very strong host galaxy contribution at 1.6\,$\mu$m,
   and starlight exceeds the dust luminosity even at wavelengths as
   long as 3.5\,$\mu$m.
   In principle, class C is similar to class B but with stronger host galaxy
   contribution. 
    
\item[D)] Three sources immediately below the dotted box in 
   Fig.\,\ref{fig_nir_mir_cc}: 
   Their SEDs can be explained by a slightly reddened AGN (similar
   to class A)
   plus an added component of starlight contributing significantly at
   1.6--3\,$\mu$m.   

\item[E)] Three sources with quasar-like SED colors (inside the dotted box in 
   Fig.\,\ref{fig_nir_mir_cc}): 
   Their SEDs overlap with the low luminosity end of the quasar SEDs. 
   In the orientation-based unified scheme, these sources could be borderline
   so that most of the dust torus is visible but 
   the broad line region and the UV-optical continuum 
   are obscured.
\end{itemize}

 \begin{figure*}[htb]
   \begin{center}
   \includegraphics[height=14.5cm, angle=90]{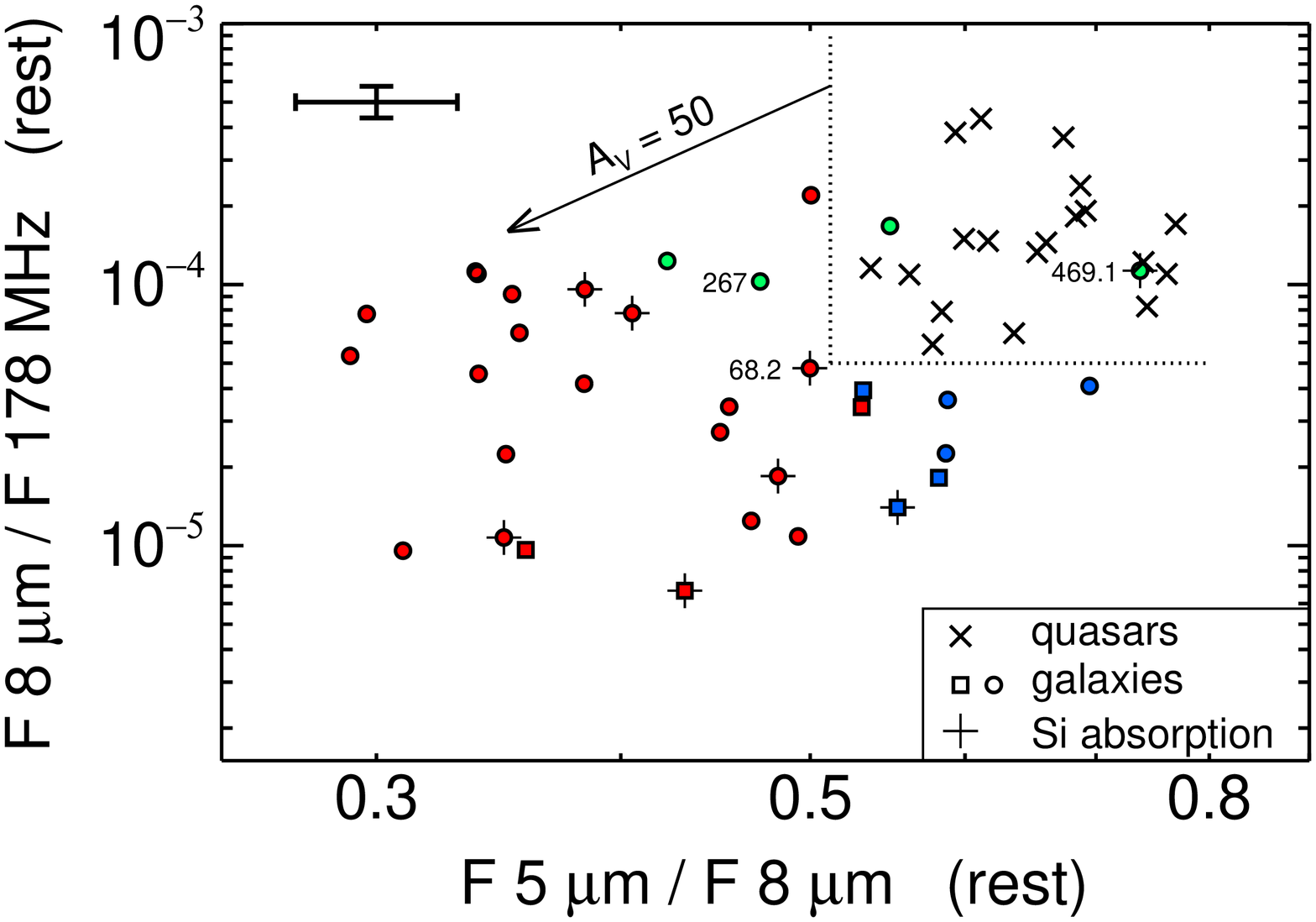}
   \caption{Infrared-radio color-color diagram.
     The dotted line marks the region occupied by quasars. 
     The color-coding and symbols of radio galaxies correspond to those in 
     Figures~\ref{fig_ir_radio_lum} and~\ref{fig_nir_mir_cc}.
     The two sources 3C\,267 and 3C\,469.1 with
     spectroscopically-detected silicate
     absorption are labeled as well as the two low-excitation radio 
     galaxies 3C\,68.1 and 3C\,469.1. 
     The error bar in the upper left corner represents an rms of
     15\%. 
     The $A_{V}$ arrow indicates screen extinction with 
     the reddening law given in \S\ref{sec_rg}.
}
   \label{fig_ir_radio_cc}
   \end{center}
 \end{figure*}
While the rest 8--10~\micron\ range is poorly sampled, eight galaxies show
declines in this range that
could be caused by silicate absorption. (The MIPS-24 filter, 50\%
transmission at 20.8--29.3\,$\mu$m, requires $z\la1.8$ for the
silicate feature to fall into its range.) 
One of these sources (3C\,469.1, $z=1.336$) has an IRS spectrum  available. 
It shows a broad silicate absorption with optical depth
$\tau_{ 9.7} \approx 0.55$
corresponding to $A_{V}\approx10$, consistent with its 
position in Fig.\,\ref{fig_nir_mir_cc}. 
This supports the view that
the SED declines in the other 
radio galaxies are due to silicate absorption, too.
The photometric silicate absorption sources show a wide range of colors
(Fig.\,\ref{fig_nir_mir_cc}), but only one galaxy (3C\,469.1) is on the blue (right)
side, where low-extinction sources reside.
The important conclusion is that the silicate feature requires
considerable extinction to be present in at least  some of the 
radio galaxies, and  this is largely
independent of the SED class.\footnote{
  The photometric silicate absorption sources are 
  3C\,68.2, 3C\,222, 3C\,249, 3C\,250, 3C\,266, 3C\,305.1, 3C\,324,
     and 3C\,469.1.
  These galaxies lie in the redshift range
  $1.08<z<1.83$, suggesting that in this 
  range the broad band 16\,$\mu$m/24\,$\mu$m filter
  combination is able to register silicate absorption, if strong
  enough. For comparison, this
  redshift range contains 20 more radio galaxies with 16 and 24~\micron\
  photometry available but without silicate
  absorption signatures in their broadband SEDs. Four of these sources 
  (3C\,13, 3C\,266, 3C\,267, 3C\,356) have IRS spectra available,
  but significant silicate absorption is
  detected only in one of them (3C\,267, $z=1.14$, $\tau_{9.7}\approx0.2$). 
  At low redshift, the detection of silicate
  absorption appears not to be directly correlated with other
  absorption signatures, perhaps because of complex geometry and/or
  varying silicate dust abundance 
  (e.g., Haas et al.\ 2005, Ogle et al.\ 2006, Cleary et al.\ 2007).
  A detailed analysis of the high-$z$ 3CR spectra and the photometric
  detectability of silicate features will be presented elsewhere
  (Leipski et al.\ in prep.).
}


If radio galaxies are misaligned quasars, as proposed in the unified scheme,
reddening of individual galaxies should be 
correlated with their extinction.
Figure~\ref{fig_ir_radio_cc} shows that this is indeed the case.
Quasars populate a distinct region of this diagram
characterized by high MIR/radio and 
blue NIR--MIR colors. 
Most radio
galaxies spread towards fainter MIR/radio and redder NIR/MIR.
Under the reasonable assumption 
that the emission at  5--8\,$\mu$m is 
not affected by the host galaxy,  
de-reddening along the direction of the extinction
vector can place each radio galaxy inside the region populated by quasars.
Thus individual radio galaxies can be explained 
as reddened quasars, consistent with the
orientation-dependent unified scheme. 


The typical amount of radio galaxy
reddening, $A_V \approx 50$ for an obscuring screen
(Fig.~\ref{fig_ir_radio_cc}), corresponds to 
a hydrogen column density $N_{H} \approx 9\times10^{22}$~cm$^{-2}$
(for $A_V = 5.6\times10^{-22}$~mag/cm$^{-2}$ --- Seward et al.\ 1999). 
This is close to but below the Compton-thick limit 
($N_{H} = 10^{24}$~cm$^{-2}$).
Screen extinction is a
simplification, and one may expect a more complex geometry.
If emitting dust particles are
spatially mixed with the absorbing ones, 
the amount of dust has to be higher for the same observed reddening, 
typically by a factor 3--5.\footnote{
The transmission factors are $\exp(-\tau_\lambda)$ and 
$(1-\exp(-\tau_\lambda))/\tau_\lambda$
for the screen and the mixed case, respectively, with 
$\tau_\lambda = 0.916\times A_{\lambda}$ 
(Disney et al.\ 1989).
}
Thus there could very well be enough gas present to render the AGN
Compton-thick. 

There is, unfortunately, no independent measurement of reddening for
individual galaxies, nor is it certain that a Galactic reddening
curve applies to AGN.
Thus it is still an open question  whether after de-reddening
there will remain a difference in the 8~$\mu$m/178~MHz ratio between
radio galaxies and quasars. 
If such a difference remains, with quasars having a higher
8~$\mu$m/178~MHz ratio than radio galaxies, 
then either our screen extinction premise is too simple or the MIR
luminosity of quasars is enhanced by 
an additional --- potentially non-thermal --- contribution.
Our {\it Chandra}  X-ray observations of a subset of
the sample will provide independent estimates of the extinction 
towards the nuclei (Wilkes et al., in prep.). 

To summarize, while quasars exhibit a uniform SED shape which can
be explained by a centrally heated dust distribution, 
radio galaxies show a diversity of SED shapes.  In all cases,
however, the radio galaxy SEDs are
consistent with being intrinsically a quasar 
modified by absorption of the dust emission 
and addition of some amount of host galaxy starlight. 

\subsection{Evolutionary effects and non-thermal contributions}

Studying powerful 3CR sources at $z<1$, Ogle et al.\ (2006) found 
evidence for a population of accretion-inefficient radio
galaxies, 
in which the jet/lobe may be powered by extraction of rotational black
hole energy. 
These sources, mainly optically-classified low-excitation radio
galaxies (LERGs), have a 
15\,$\mu$m luminosity below 8\,$\times$\,10$^{\rm 43}$\,erg/s 
and a 
luminosity ratio $L_{\rm 15\mu m}$\,/\,L$_{\rm 178MHz}$\,$<$\,10. 
In contrast, with the reaonable assumption that $L_{\rm 8\mu m}\la L_{\rm 15\mu m}$, 
all our $z>1$ sources have observed MIR luminosity 
$L_{\rm 8\mu m}>5\times10^{\rm 44}$\,erg/s, 
which is expected to be even higher after de-reddening. 
Also, the two LERGs (3C\,68.2, 3C\,469.1) in our sample show a high 
luminosity ratio L$_{\rm 8\mu m}$\,/\,L$_{\rm 178MHz}$\,$>$\,10 
comparable to that of quasars
(Fig.\,\ref{fig_ir_radio_lum}).  
From this, our data do not support the existence of
an accretion-inefficient population 
among the powerful 3CR sources at $z>1$. 
A possible explanation for the deficit of 
optical high-excitation line luminosity (for instance
[\ion{O}{3}]~$\lambda$5007\,\AA)
in our two LERGs may be extinction of the narrow-line region. 
On the other hand, some of our radio galaxies with very strong
host contribution (plotted as squares in Fig.\,\ref{fig_ir_radio_cc})
are expected after de-reddening to lie at the low end of the 
$L_{\rm 8\mu m}/L_{\rm 178MHz}$ distribution. Hence compared
with the strength of both the host and the radio lobes, these galaxies 
are relatively weak in the MIR and may represent a
population at the beginning of a different evolutionary state. 

Some authors have attributed the excess emission of quasars compared
to radio galaxies to nonthermal emission from synchrotron jets.
For example, Cleary et al.\ (2007) fitted the SEDs and
spectra of 3CR sources at $0.5<z<1$ 
with a combination of a spherically symmetric dust model and a 
jet+lobe synchrotron component.
They attributed half of the factor of four excess in the 15\,$\mu$m luminosities
of steep-spectrum quasars relative to radio galaxies to
a non-thermal component. 
If such a non-thermal component were present
in our 3CR sources at $z>1$, 
it would show up in Fig.\,\ref{fig_ir_radio_cc} as an offset by
about a factor of two between dereddened radio galaxies and quasars.
This conclusion is, however, dependent on both the reddening law and
on the radiative transfer and thus the geometry of the emitting region.  
In order to draw definite conclusions 
about any MIR luminosity excess,
detailed  radiative transfer modelling 
is required
(Heymann et al., in prep.).
Spherically symmetric models are wholly inadequate for this purpose.
In an inclined disk-like system, some fraction of the MIR emission is
likely to have very little obscuration while 
the bulk of the MIR emission is heavily obscured, and 
no spherical model 
can properly account for this  geometry.  All we can say at the
moment is that our data appear consistent with a simple thermal
interpretation and show no evidence for a non-thermal component.

\section{Conclusions}

The 3CR sample at $1<z<2.5$ represents the most luminous
steep-spectrum quasars (type~1 AGN) and powerful double-lobed radio
galaxies (type~2 AGN).  This sample is nearly unbiased by
orientation.  We have defined subsamples of 19 quasars and 33 radio
galaxies matched in isotropic rest 178\,MHz luminosity and have
obtained {\it Spitzer} 3.6--24\,$\mu$m photometry.  The main results
are:

\begin{itemize}
\item[1)] Quasars all have  similar energy distributions in the rest
   frame  1.6--10\,$\mu$m range, and their ratio of MIR to radio
   luminosity is also nearly constant.  This is consistent with results
   seen previously in lower-redshift samples.

\item[2)]   The rest frame 1.6--10\,$\mu$m SEDs of radio galaxies can  be
   explained as reddened quasars, consistent with
   orientation-dependent unification. 
   Various amounts of extinction of the AGN emission combined
   with addition of host galaxy starlight can
   explain the diversity of radio galaxy SEDs. 
   
\item[3)] If the extinction is sufficiently large, 
  there is no need to invoke a beamed synchrotron contribution to
  explain the MIR luminosity difference between quasars and radio
  galaxies. 
  The actual amount of extinction has to be derived from additional
  observations. 

\item[4)] 
  The above results hold also for splitting our sample in redshift and
  luminosity; 
  within our sample we do not find any trends with redshift or luminosity.  

\item[5)] 
  At rest frame 8\,$\mu$m, quasars are 3 times more luminous than radio galaxies. 
   If this difference applies also
   to high-redshift, radio-quiet AGN, then MIR (24\,$\mu$m) 
   surveys are strongly biased in favour of type-1 and against type-2
   AGN. This will make it very difficult to resolve the AGN/starburst
   degeneracy with only broadband
   SEDs, and spectral line diagnosis will be required.  

\end{itemize}

While our near-mid-IR SEDs provide a fundamental set of high
luminosity AGN templates, we expect to derive 
more stringent conclusions from proper two-dimensional
radiative transfer modelling in combination with  {\it Spitzer}
MIR spectra, {\it Chandra} X-ray observations, and
{\it Herschel} far-IR/sub-mm data.




\acknowledgments
{\noindent \it Acknowledgments:}
This work is based on observations made with the Spitzer
  Space Telescope, which is operated by the Jet Propulsion Laboratory,
  California Institute of Technology under a contract with NASA.
  Support for this work was provided by NASA through an award issued by
  JPL/Caltech.  This research has made use of the NASA/IPAC
  Extragalactic Database (NED) which is operated by the Jet Propulsion
  Laboratory, California Institute of Technology, under contract with
  the National Aeronautics and Space Administration. 
  M.H.\ is supported by the 
  Nordrhein--Westf\"alische Akademie der Wissenschaften.



{\it Facilities:} \facility{Spitzer}.



\end{document}